\RequirePackage{fix-cm}
\documentclass[smallextended]{svjour3}       
\smartqed  
\usepackage{graphicx}
\graphicspath{{screenshots/}}
\usepackage{hyperref}
\usepackage{amscd, amssymb, mathtools}
\usepackage{float}
\usepackage{caption}
\usepackage{array}

\usepackage{csquotes}
 \journalname{Journal of Automated Reasoning}
\begin{document}

\title{Certified Quantum Computation in Isabelle/HOL\thanks{Work on this paper was supported by the European Research Council Advanced Grant ALEXANDRIA (Project 742178).}
}

\titlerunning{Certified Quantum Computation in Isabelle}        

\author{Anthony Bordg \and Hanna Lachnitt \and Yijun He}

\authorrunning{A. Bordg \and H. Lachnitt \and Y. He} 

\institute{Anthony Bordg \at
              Department of Computer Science and Technology, University of Cambridge, Cambridge, UK \\
              \email{apdb3@cam.ac.uk}           
           \and
           Hanna Lachnitt \at Computer Science Department, Stanford University, Stanford, US \\
           \email{lachnitt@stanford.edu}
           \and 
           Yijun He, University of Cambridge, Cambridge, UK 
}

\date{Received: date / Accepted: date}

\maketitle

\begin{abstract}
In this article we present an ongoing effort to formalise quantum algorithms and results in quantum information theory using the proof assistant Isabelle/HOL. Formal methods being critical for the safety and security of algorithms and protocols, we foresee their widespread use for quantum computing in the future. We have developed a large library for quantum computing in Isabelle based on a matrix representation for quantum circuits, successfully formalising the no-cloning theorem, quantum teleportation, Deutsch's algorithm, the Deutsch-Jozsa algorithm and the quantum Prisoner's Dilemma. We discuss the design choices made and report on an outcome of our work in the field of quantum game theory.
\keywords{Isabelle/HOL \and Certification \and Quantum computing \and No-cloning \and Quantum teleportation \and Deutsch's algorithm \and Deutsch-Jozsa algorithm \and Quantum Prisoner's Dilemma}
 \subclass{03B35 \and 03B15 \and 81P68 \and 68Q12}
\end{abstract}

\section{Introduction}
\label{intro}
On January 4th 2017 the computer scientist L\'aszl\'o Babai retracted a claim he made in a preprint\footnote{\url{https://arxiv.org/abs/1512.03547}} after the mathematician Andr\'es Helfgott spotted an error in his work. Back in 2015, Babai's result, dealing with the so-called graph isomorphism problem, a central problem in the field of computer algorithms, was dubbed ``the theoretical computer science advance of the decade''. In the meantime Helfgott had spent months reviewing Babai's algorithm in preparation for a talk at the Bourbaki seminar to report on Babai's major result. On January 9th 2017 in a new (last?) unexpected twist, Babai announced a fix for his error and he restored his claim. It seems that Helfgott is confident that the fix is correct, but Babai's paper is still unpublished as of September 23, 2020. \\
This story exemplifies the difficulty making sure that an algorithm obeys its specification. This challenge might be even harder with respect to quantum algorithms, since our intuition weakens when one moves from the classical world to the quantum realm. Fortunately, formal methods can help with the task of certifying that a quantum algorithm obeys its specification. Recent work in that direction include the formalisation of Grover's algorithm in Isabelle by Liu {\em et al.}\ \cite{10.1007/978-3-030-25543-5_12}. However, for their formalisation of Grover's algorithm  the authors use a tailored quantum Hoare logic. One should also mention the QWIRE project by Rand {\em et al.}\ \cite{2018arXiv180300699R} for quantum circuits, although the authors use a different approach since their work is an embedding of the QWIRE quantum circuit language in the proof assistant Coq to formally prove properties of those circuits. The closest to the present work is the work of Boender {\em et al.}\ \cite{Boender2015FormalizationOQ} which culminates in the formalisation of the quantum teleportation protocol using the proof assistant Coq, this algorithm becoming the de facto benchmark in the field. This benchmark is successfully reached and surpassed in our work. \\
In this paper we present a large formalisation of results in quantum computation and quantum information theory developed in the proof assistant Isabelle. Our library\footnote{{\em Isabelle Marries Dirac}, freely available on GitHub \url{https://github.com/AnthonyBordg/Isabelle_marries_Dirac}.} includes the quantum teleportation protocol, the no-cloning theorem, Deutsch's algorithm, the Deutsch-Jozsa algorithm and the quantum Prisoner's Dilemma. To the best of our knowledge a formalisation of these last four classic results has never been done before. We start with the basics of quantum computing in Sections \ref{sec:qubits}, \ref{sec:gates} and \ref{sec:meas}. We then introduce the aforementioned results formalised in the library in Section \ref{sec:algo}. Throughout the article we discuss the design choices made. Finally, we outline an unexpected outcome of our formalisation in Section \ref{ssec:qPD}. 

\section{Qubits and Quantum States}
\label{sec:qubits}

In the classical model of computation the bit is the fundamental unit of information. There are two classical states for a bit, namely $0$ and $1$. In quantum computing the bit is superseded by the quantum bit, the so-called {\em qubit}, which becomes the fundamental unit of information. In the same way, any qubit has a (quantum) state, but the situation is more involved. \\
For the sake of simplicity, let us start with a 1-qubit system. In that case, the quantum state of our qubit is a normalised vector in a 2-dimensional complex vector space. Using the Dirac notation introduced in quantum mechanics, a column vector in that space is denoted by $|\psi\rangle$, where $\psi$ is a mere label, and the vector $|\psi\rangle$ is called a {\em ket}. In that context the two elements of the computational basis, namely 
\[
\begin{pmatrix} 
1 \\ 0
\end{pmatrix} \;,
\begin{pmatrix} 0 \\ 1
\end{pmatrix}\;,
\]
are denoted by $|0\rangle$ and $|1\rangle$, respectively. Actually, the label for the $n$th element of the computational basis corresponds to the binary expression of $n$, hence $|0\rangle$ should not be confused with the zero vector, namely 
\[
\begin{pmatrix} 0 \\ 0 \end{pmatrix} \;.
\]
The zero vector being not normalised, it is not a quantum state. \\
So, in the computational basis a state of our qubit is a linear combination $\alpha_0\, |0\rangle + \alpha_1\, |1\rangle$ such that $\alpha_0$ and $\alpha_1$ are complex numbers satisfying the normalisation constraint 
\[
\vert \alpha_0\rvert^2 + \lvert \alpha_1 \rvert^2 = 1.
\] 
In the quantum world, the coefficients $\alpha_0$ and $\alpha_1$ are called {\em amplitudes}, and one sometimes uses the word {\em superposition} instead of the phrase {\em linear combination}. \\
For a 2-qubit system the state of a qubit becomes a normalised vector in a 4-dimensional complex vector space. If $|00\rangle, |01\rangle, |10\rangle, |11\rangle$ denote the elements of the computational basis, then such a state is a superposition 
\[
\alpha_{00}\, |00\rangle + \alpha_{01}\, |01\rangle + \alpha_{10}\, |10\rangle + \alpha_{11}\, |11\rangle\;,
\]
with $\lvert \alpha_{00} \rvert^2 + \lvert \alpha_{01} \rvert^2 + \lvert \alpha_{10}\rvert^2 + \lvert\alpha_{11}\rvert^2 = 1$. \\
In order to model the quantum states of qubits we exploit Isabelle's system module for dealing with a hierarchy of parametric theories, the so-called {\em locales} \cite{Ballarin2014}.
In our library the locale {\em state} provides the context for talking about the quantum states of a $n$-qubit system.

\begin{figure}[H]
	\includegraphics[scale=0.32]{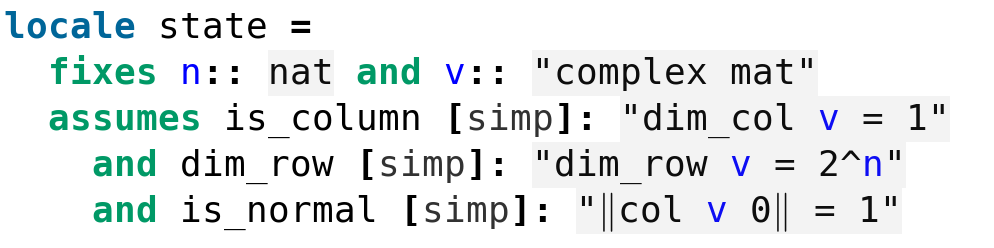}
	 \caption{The locale {\em state} in {\em Quantum.thy}}
\end{figure} 
\noindent
In this locale $v$ is a complex matrix, but the condition {\em is\_column} ensures it is a column matrix, {\em i.e.} a column vector. We choose to model quantum states by column matrices instead of vectors, since this design choice will come in handy when applying quantum gates to quantum states (more on that later). The condition {\em dim\_row} relates to the dimension of the ambient vector space and the condition {\em is\_normal} provides the normalisation constraint. We also introduce the corresponding set of quantum states of a given dimension, where this time we can directly use complex vectors.

\begin{figure}[H]
	\includegraphics[scale=0.32]{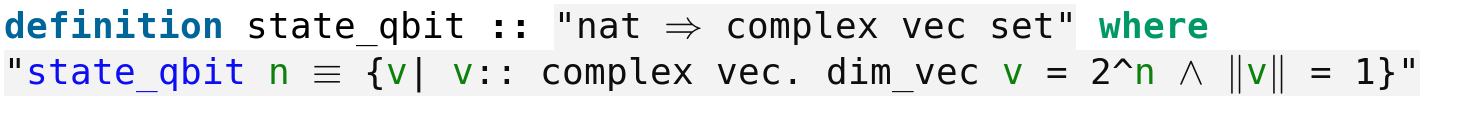}
	\caption{The definition of the set {\em state\_qbit} in {\em Quantum.thy}}
\end{figure}
\noindent
Of course, given the context provided by the locale {\em state} we can prove that $v$ (or rather the first column of $v$) belongs to {\em state\_qbit n}.

\begin{figure}[H]
	\includegraphics[scale=0.32]{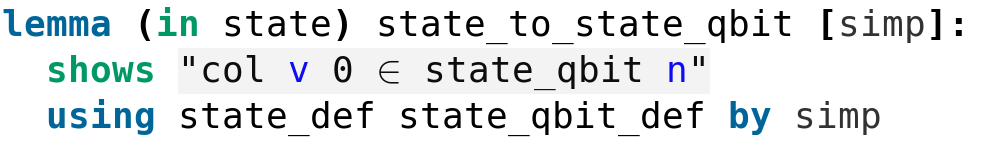}
\end{figure}
\noindent
One can also go the other way around, {\em i.e.}\ from vectors to column matrices. We take this opportunity to introduce in our library Dirac's ket notation, since Isabelle allows some syntactic sugar.

\begin{figure}[H]
	\includegraphics[scale=0.32]{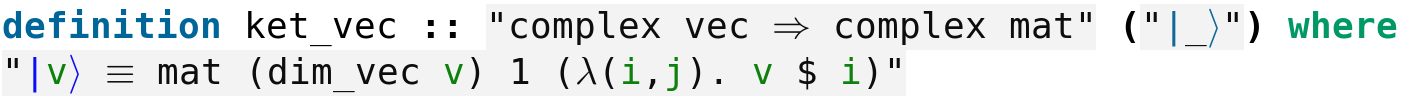}
	\caption{Dirac's ket notation}
\end{figure}
\noindent
In our library we implicitly work in the computational basis, hence the amplitudes have to be understood accordingly.

\section{Quantum Gates}
\label{sec:gates}	

Like their classical counterparts, quantum gates are used to manipulate information. More exactly, quantum gates are ways of manipulating the quantum states of qubits. Usually there are two kinds of representations for quantum gates, namely circuit representations and matrix representations. Since it is not possible to directly work in Isabelle with circuits, we choose in that context the more convenient matrix representations. In this process we take advantage of the nice library for matrices developed by Thiemann and Yamada \cite{Jordan_Normal_Form-AFP}.  Then, a quantum state $|\psi\rangle$ being in particular a column matrix, the action of a quantum gate $U$ on $|\psi\rangle$ is simply given by the matrix multiplication $U \,|\psi\rangle$, denoted $U \,^*\, |\psi\rangle$ in Isabelle. \\
However, not every matrix is a quantum gate. Quantum gates belong to a specific class of complex matrices. Actually, given a $n$-qubit system the quantum gates are exactly the $2^n\times 2^n$ matrices that are unitary. In order to explain what {\em unitary} means, we need to introduce the Hermitian conjugate of a complex matrix. Let $U$ be a complex matrix, its Hermitian conjugate $U^\dagger$ is the complex conjugate of its transpose, namely $(U^t)^*$. In different contexts people use different notations for the Hermitian conjugate, but the dagger operator is commonly used in quantum mechanics, and we keep this notation in the library. A complex square matrix $U$ is said to be unitary if $U^\dagger \,U = U\, U^\dagger = I$, {\em i.e.}\ if its inverse is given by its Hermitian conjugate.	

\begin{figure}[H]
	\includegraphics[scale=0.32]{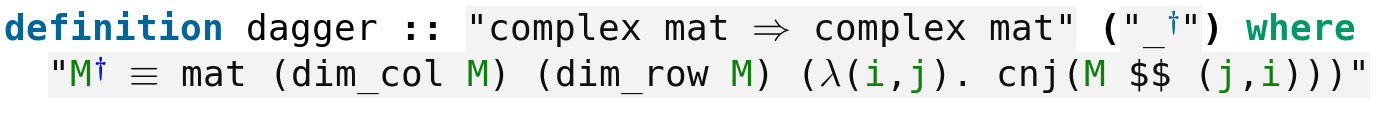}
	\includegraphics[scale=0.32]{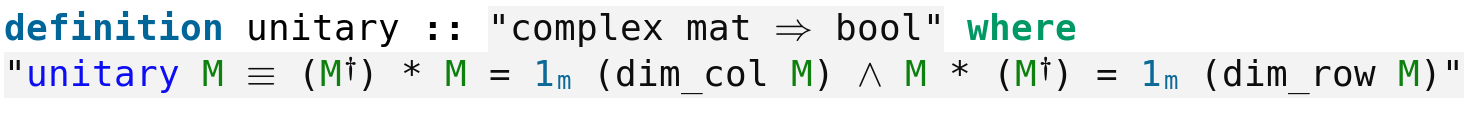}
	\caption{The definitions of the Hermitian conjugate of a matrix and the {\em unitary} predicate, respectively ({see~\em Quantum.thy})}
\end{figure}
\noindent
In Isabelle we encapsulate the definition of a quantum gate inside a dedicated locale.

\begin{figure}[H]
	\includegraphics[scale=0.32]{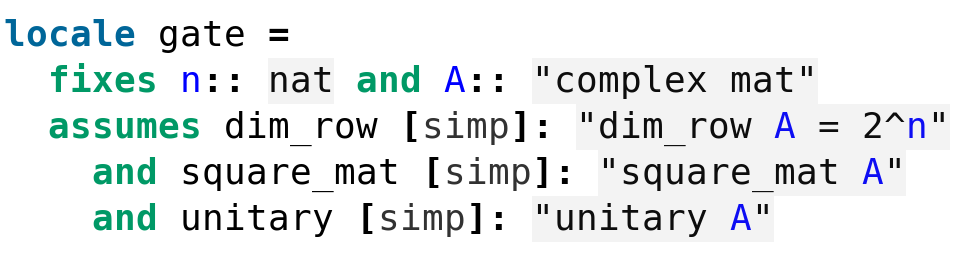}
	\caption{The locale {\em gate} in {\em Quantum.thy}}
\end{figure}
\noindent
What is the idea behind unitarity? Unitary matrices are length-preserving. One has $\lVert U \, |v\rangle\rVert = \lVert |v\rangle \rVert$ for every unitary matrix $U$ and every ket $|v\rangle$ such that their multiplication is well defined. Given the normalisation constraint in quantum states, it is no wonder quantum gates should be unitary matrices.

\begin{figure}[H]
	\includegraphics[scale=0.32]{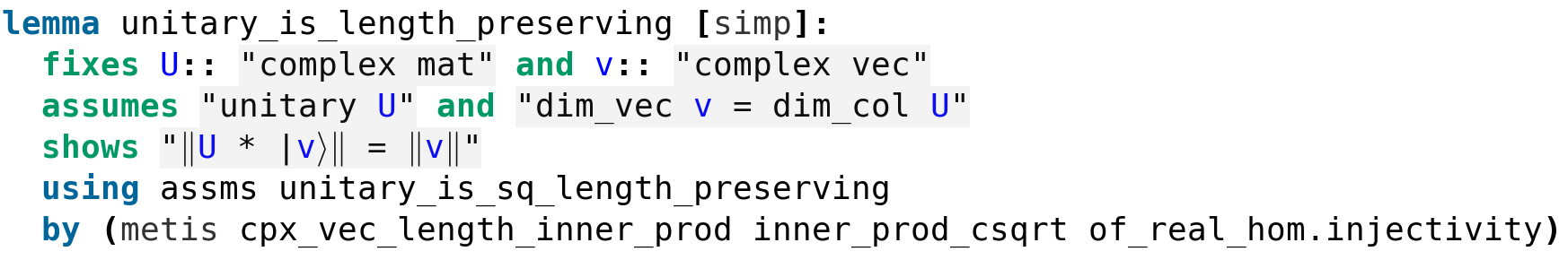}
	\caption{The statement and the proof that unitary matrices preserve length}
\end{figure}
\noindent
Actually, unitary matrices are the only matrices that preserve length. To prove this result one needs the following key lemma.

\[
(M\,|\psi\rangle)^\dagger = \langle\psi|\,M^\dagger,
\] 
where $\langle\psi|$ is Dirac's {\em bra} notation, namely if $|\psi\rangle$ is the column vector

\[
\begin{pmatrix}a_1 \\ \vdots \\ a_{2^n} \end{pmatrix},
\]	
then its bra is the corresponding row vector with conjugate coefficients 
\[
\begin{pmatrix} 
a_1^* \cdots a_{2^n}^*
\end{pmatrix}.
\]

\begin{figure}[H]
	\includegraphics[scale=0.32]{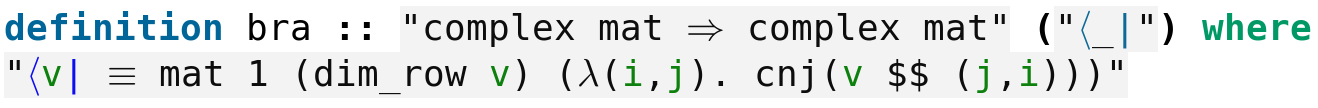}
	\includegraphics[scale=0.32]{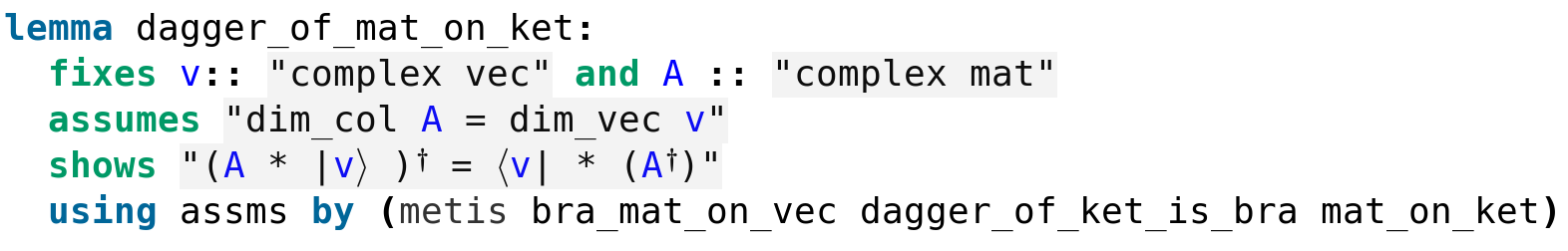}
	\caption{A key lemma to prove that length-preserving matrices are unitary}
\end{figure}
\noindent
Using this lemma and the many results on the dagger operator and unitary matrices provided in the library, one can eventually prove that length-preserving matrices are unitary.

\begin{figure}[H]
	\includegraphics[scale=0.32]{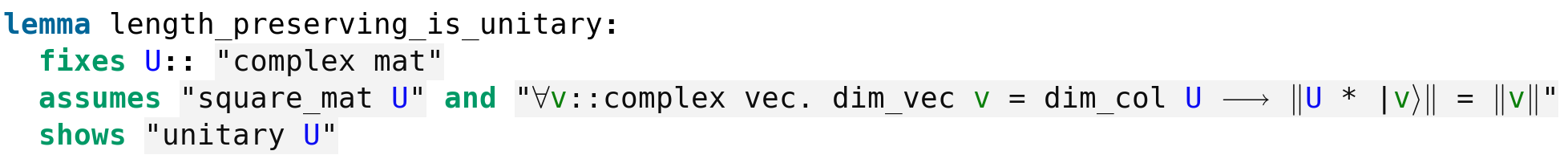}
	\caption{Length-preserving matrices are unitary}
\end{figure}
\noindent
Now, we introduce our first quantum gate, namely the Hadamard gate $H$. It is a single-qubit gate, {\em i.e.}\ a $2 \times 2$ unitary matrix,

\[
H = \frac{1}{\sqrt{2}}
\begin{pmatrix}
1 & 1 \\
1 & -1
\end{pmatrix}.
\]

\begin{figure}[H]
	\includegraphics[scale=0.32]{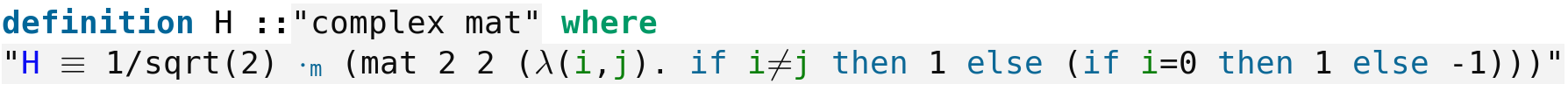}
	\caption{The definition of the Hadamard gate in {\em Quantum.thy}}
\end{figure}
\noindent
One can easily check that $H$ is unitary and self-adjoint, {\em i.e.}\ $H^\dagger = H$. 

\begin{figure}[H]
	\includegraphics[scale=0.32]{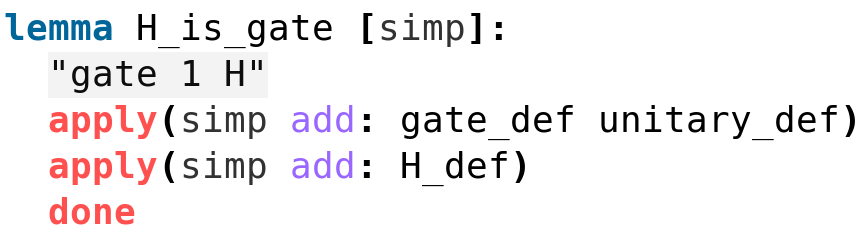} \\
	\includegraphics[scale=0.32]{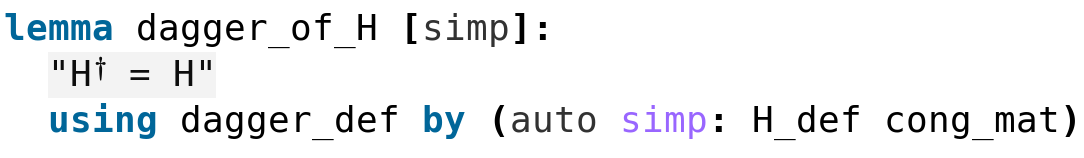}
	\caption{The formal proof in Isabelle that $H$ is a gate and is self-adjoint}
\end{figure}
\noindent
The action of $H$ on the basis elements is given as follows. 

\begin{gather*}
|0\rangle \mapsto \frac{1}{\sqrt{2}} \,(|0\rangle + |1\rangle) \\
|1\rangle \mapsto \frac{1}{\sqrt{2}} \,(|0\rangle - |1\rangle)
\end{gather*}	
As a consequence, $H$ maps a state $\alpha_0\,|0\rangle + \alpha_1\,|1\rangle$ to $\frac{\alpha_0 + \alpha_1}{\sqrt{2}}\,|0\rangle + \frac{\alpha_0 - \alpha_1}{\sqrt{2}}\,|1\rangle$. \\		
Since $H$ creates a superposition, it is truly a quantum gate. Maybe somewhat puzzling for the beginner is the fact that $H$ is sometimes described as a ``square-root of NOT'' gate.  One simply means that it turns $|0\rangle$ ({\em resp.}\ $|1\rangle$) into $\frac{1}{\sqrt{2}} \,(|0\rangle + |1\rangle)$ ({\em resp.}\ $\frac{1}{\sqrt{2}} \,(|0\rangle - |1\rangle)$), so ``halfway''  between $|0\rangle$ and $|1\rangle$. \\
Before introducing our first example of a 2-qubit gate, we need to say a few words on the initial states of a 2-qubit system. Actually, such states are given by the tensor product of the states of each qubit. For instance, if the first qubit is in the initial state $|0\rangle = \begin{psmallmatrix} 1 \\ 0 \end{psmallmatrix}$ and the second one is in the initial state $|1\rangle = \begin{psmallmatrix} 0 \\ 1 \end{psmallmatrix}$, then the initial state of the combined system is

\begin{equation*} \label{eq1}
\begin{split}
|0\rangle \otimes |1\rangle  & =  \begin{psmallmatrix} 1 \\ 0 \end{psmallmatrix} \otimes \begin{psmallmatrix} 0 \\ 1 \end{psmallmatrix} \\
& = \begin{psmallmatrix} 0 \\ 1 \\ 0 \\ 0 \end{psmallmatrix} \\
& = |01\rangle.
\end{split}
\end{equation*}
Now, an interesting 2-qubit quantum gate is the controlled-NOT gate (cNOT). Its matrix representation is given by

\[
cNOT =
\begin{pmatrix}
1 & 0 & 0 & 0 \\
0 & 1 & 0 & 0 \\
0 & 0 & 0 & 1 \\
0  & 0 & 1 & 0
\end{pmatrix},
\]
and one easily checks that the cNOT gate is unitary and it is again self-adjoint.

\begin{figure}[H]
	\includegraphics[scale=0.32]{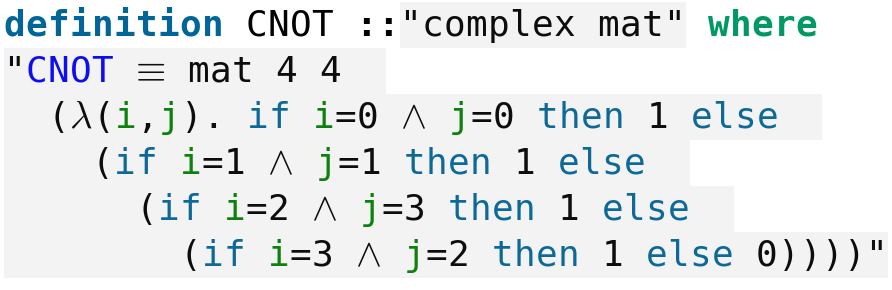} \\
	\includegraphics[scale=0.32]{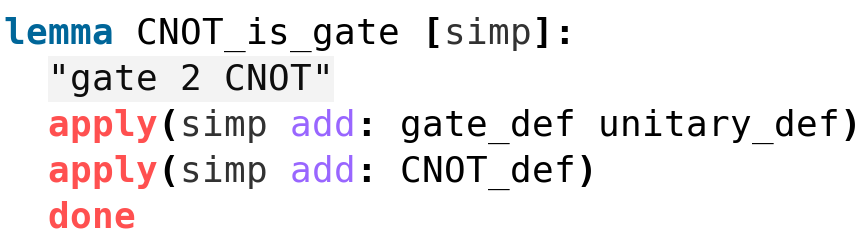}
	\caption{The formal proof that cNOT is a 2-qubit gate}
\end{figure}
\noindent
The cNOT gate maps the basis elements $|00\rangle, |01\rangle, |10\rangle, |11\rangle$ to $|00\rangle, |01\rangle, |11\rangle, |10\rangle$, respectively. In other words, the cNOT gate flips the second qubit (the so-called target qubit) if the first qubit (the so-called control qubit) is $1$ and does nothing otherwise. One summarizes the action of the cNOT gate with the following handy piece of notation
\[
|xy\rangle \mapsto |x\;x\oplus y\rangle,
\]
where $\oplus$ denotes the addition modulo 2. \\
The cNOT gate can be used to perform non-classical computations. For instance, starting with the $|00\rangle$ state and applying a Hadamard gate to the first qubit followed by a cNOT gate, one creates the state $\frac{1}{\sqrt{2}}\,(|00\rangle + |11\rangle)$, which is a highly non-classical state, a so-called Bell's state (more on that later). \\
To put everything together, let us assume that we have a 3-qubit system. Moreover, assume that we want to apply an Hadamard gate to the first qubit and a cNOT gate to the second and third qubits. The initial state of the combined system is given by $|x\rangle \otimes |y\rangle \otimes |z\rangle$ which is a 8-dimensional column vector with $|x\rangle$ ({\em resp.}\ $|y\rangle$, $|z\rangle$) denoting the initial state of the first ({\em resp.}\ second, third) qubit. Since the tensor product is associative, we omit the parentheses in $|x\rangle \otimes |y\rangle \otimes |z\rangle$. Then one can sum up the two gate applications using only one $8\times 8$ matrix, namely $H \otimes cNOT$, where $\otimes$ denotes the {\em Kronecker product} between two matrices. With this in mind we needed to formalize the Kronecker product in our library and proved that the Kronecker product of two gates is a gate as shown in the snippet of code below. This essentially amounts to  proving that the Kronecker product of two unitary matrices is a unitary matrix.

\begin{figure}[H]
	\includegraphics[scale=0.32]{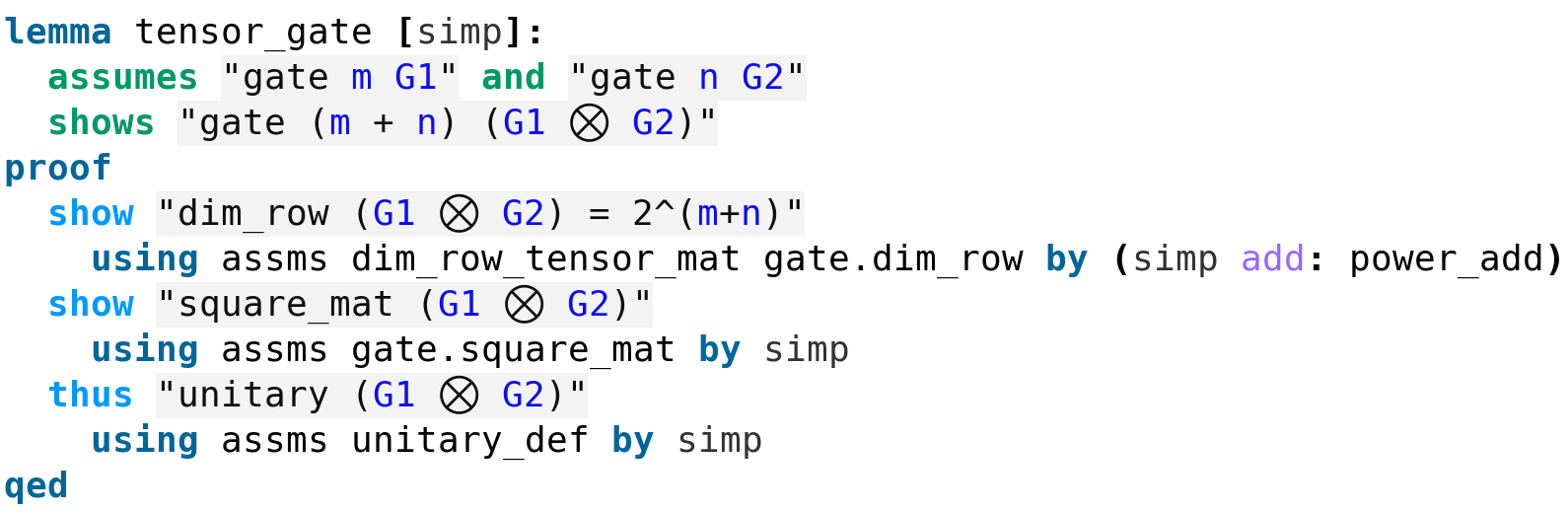}
	\caption{The Kronecker product of two quantum gates is a quantum gate (see {\em More\_Tensor.thy})}
\end{figure}
\noindent
At that point we faced a design choice connected to the important issue of legacy code in formal mathematics. Indeed, there is already a formalisation of the Kronecker product in \cite{Matrix_Tensor-AFP} but for a legacy notion of matrix which is not the one developed in \cite{Jordan_Normal_Form-AFP} and used in our library. So, we could either restart the formalisation of the Kronecker product from scratch or we could build a bridge between the two formalisations of matrices available and reuse as much as possible the code in \cite{Matrix_Tensor-AFP}. We chose the latter, using the code available as a convenient scaffolding ({\em cf.}\ our theory {\em Tensor.thy}). This choice may ease in the future the reuse of formalisations based on legacy matrices. \\
We come back to the state $\frac{1}{\sqrt{2}}\,(|00\rangle + |11\rangle)$ obtained as the result of the application of the Hadamard gate followed by the cNOT gate to the state $|00\rangle$. Actually, this state is part of a set of four states known as the {\em Bell's states} or sometimes the {\em EPR states} (EPR stands for Einstein, Podolsky and Rosen):

\begin{align*}
|\beta_{00}\rangle & = \frac{1}{\sqrt{2}}\,(|00\rangle + |11\rangle) \\
|\beta_{01}\rangle & = \frac{1}{\sqrt{2}}\,(|01\rangle + |10\rangle) \\
|\beta_{10}\rangle & = \frac{1}{\sqrt{2}}\,(|00\rangle - |11\rangle) \\
|\beta_{11}\rangle & = \frac{1}{\sqrt{2}}\,(|01\rangle - |10\rangle)\;.
\end{align*}

\begin{figure}[H]
	\includegraphics[scale=0.32]{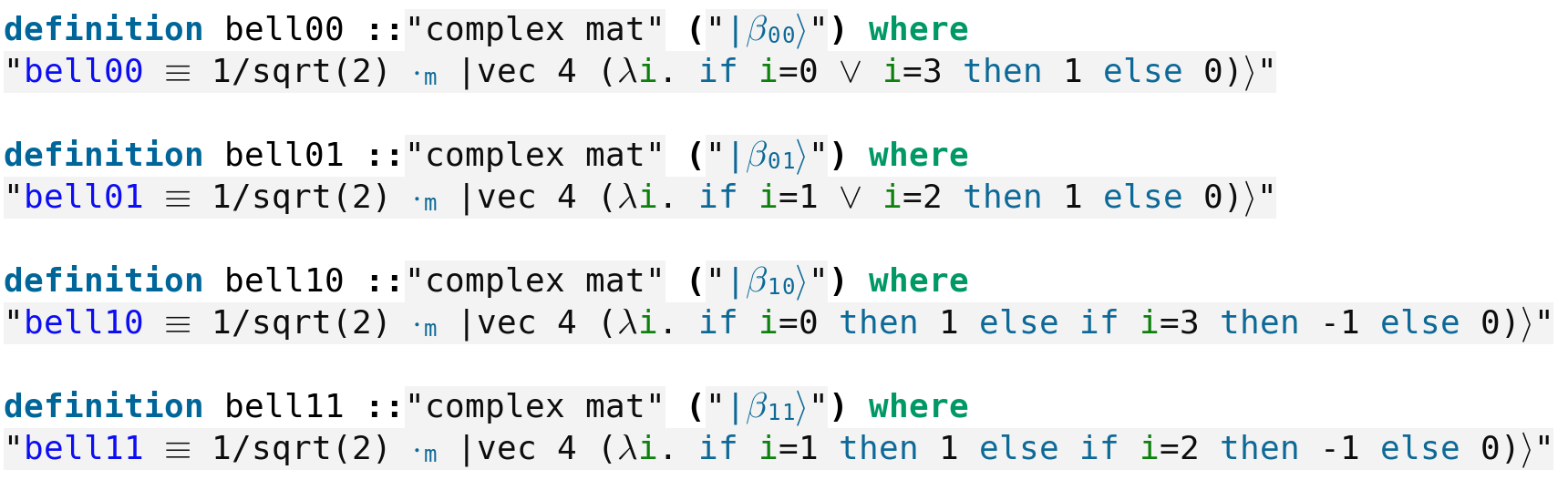}
	\caption{The Bell's states (see {\em Quantum.thy})}
\end{figure}
\noindent
The peculiarity of these states resides in the fact that they cannot be written as the tensor product of two 1-qubit states. These states are said to be {\em entangled}. {\em Entanglement}, one of the key concepts in quantum mechanics, is simply the fact that not every state is a tensor product of smaller states.

\begin{figure}[H]
	\includegraphics[scale=0.32]{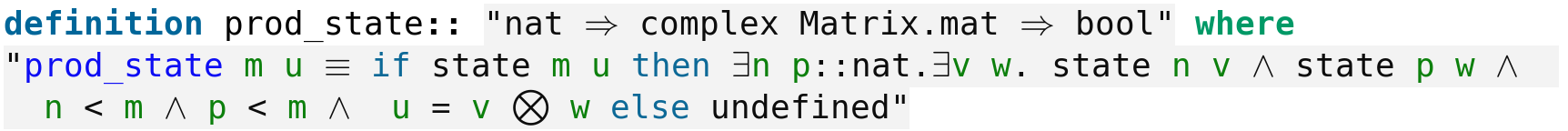}
	\includegraphics[scale=0.32]{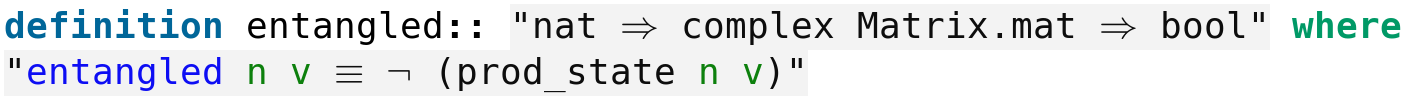}
	\caption{The property of being entangled (see {\em Entanglement.thy})}
\end{figure}
\noindent
In the case of the Bell's state $|\beta_{00}\rangle$ for instance it is very easy to prove that it cannot be written as $(\alpha_{0}|0\rangle + \alpha_{1}|1\rangle) \otimes (\alpha_{0}'|0\rangle + \alpha_{1}'|1\rangle)$ using the distributivity of the tensor product.

\begin{figure}[H]
	\includegraphics[scale=0.32]{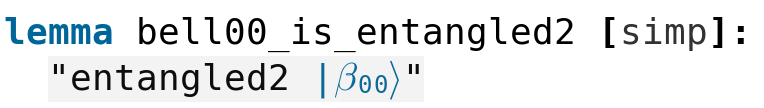}
	\caption{The proof that $|\beta_{00}\rangle$ is entangled (see {\em Entanglement.thy})}
\end{figure}
\noindent
Finally, in the library many other quantum gates are introduced like the Pauli matrices $X$, $Y$ and $Z$, the phase gate $S$, the $T$ gate\footnote{See the theory {\em Quantum} of the library, especially the subsection entitled A {\em Few Well-Known Quantum Gates}.}.

\section{Measurements}
\label{sec:meas}

Given a $n$-qubit system, the state of a qubit involves  $2^n$ amplitudes linked by a normalisation constraint. Can we determine those amplitudes? For instance take $n=1$, a quantum state has the form $\alpha_0 |0\rangle + \alpha_1 |1\rangle$ with $\lvert \alpha_0 \rvert^2 + \vert \alpha_1 \rvert^2 = 1$. Can we determine $\alpha_0$ and $\alpha_1$? The answer is no. A quantum state cannot be directly observed and the amplitudes cannot be directly determined. \\
Actually, the outcome of any measurement of our qubit through an apparatus is a classical bit of information. Moreover, the measurement will disturb the state of the qubit. Indeed, the outcome will be either $0$ with probability $\lvert \alpha_0 \rvert^2$ or $1$ with probability $\lvert \alpha_1 \rvert^2$. The sum of the probabilities should be $1$, hence the normalisation constraint of quantum states. Moreover, if the outcome happens to be $0$ ({\em resp.}\ $1$), then the post-measurement state is $|0\rangle$ ({\em resp.}\ $|1\rangle$) and the amplitudes vanish. \\
The generalisation to a multiple qubits system is straightforward. For a system of two qubits, assuming the state of our system is 
\[
\alpha_{00} |00\rangle + \alpha_{01} |01\rangle + \alpha_{10} |10\rangle + \alpha_{11} |11\rangle\;,
\]
one has 
\begin{align*}
pr(00) & = \lvert \alpha_{00} \rvert^2 \\
pr(01) & = \lvert \alpha_{01} \rvert^2 \\
pr(10) & = \lvert \alpha_{10} \rvert^2 \\
pr(11) & = \lvert \alpha_{11} \rvert^2,
\end{align*}
where pr($00$) ({\em resp.}\ pr($01$), pr($10$), pr($11$)) denotes the probability of the outcome being $0$ for both qubits ({\em resp.}\ $0$ for the first one and $1$ for the second one, $1$ for the first one and $0$ for the second one, $1$ for both qubits), and the post-measurement state is $|00\rangle$ ({\em resp.}\ $|01\rangle$, $|10\rangle$, $|11\rangle$).  \\
Now, what does happen if one has a 2-qubit system and one makes a partial measurement, {\em i.e.}\ one measures the first qubit for instance (but not the second one)? What are the probabilities pr($0$) and pr($1$) of the outcome being $0$ and $1$, respectively? One has

\begin{align*}
pr(0) & = pr(00) + pr(01) = \lvert \alpha_{00}\rvert^2 + \lvert \alpha_{01} \rvert^2 \\
pr(1) & = pr(10) + pr(11) = \lvert \alpha_{10}\rvert^2 + \lvert \alpha_{11}\rvert^2.
\end{align*}
In other words, we sum over the probabilities of measuring the whole system and getting $0$ ({\em resp.}\ $1$) for the first qubit. What is the post-measurement state of the system? To get the answer we first rewrite 
\[
\alpha_{00} |00\rangle + \alpha_{01} |01\rangle + \alpha_{10} |10\rangle + \alpha_{11} |11\rangle
\]
as
\[
|0\rangle \otimes (\alpha_{00} |0\rangle + \alpha_{01} |1\rangle) + |1\rangle \otimes (\alpha_{10} |0\rangle + \alpha_{11} |1\rangle).
\]
If the outcome of measuring only the first qubit happens to be $0$ ({\em resp.}\ $1$), then the post-measurement state of the system is 
\[
|0\rangle \otimes \frac{\alpha_{00} |0\rangle + \alpha_{01} |1\rangle}{\sqrt{\lvert \alpha_{00} \rvert^2 + \lvert \alpha_{01} \rvert^2}} \quad
(\mathit{resp.}\ |1\rangle \otimes \frac{\alpha_{10} |0\rangle + \alpha_{11} |1\rangle}{\sqrt{\lvert \alpha_{10} \rvert^2 + \lvert \alpha_{11} \rvert^2}})\,.
\]
In particular, the state of the second qubit after measuring $0$ ({\em resp.}\ $1$) for the first qubit is 
\[
\frac{\alpha_{00} |0\rangle + \alpha_{01} |1\rangle}{\sqrt{\lvert \alpha_{00} \rvert^2 + \lvert \alpha_{01} \rvert^2}} \quad
(\mathit{resp.}\ \frac{\alpha_{10} |0\rangle + \alpha_{11} |1\rangle}{\sqrt{\lvert \alpha_{10} \rvert^2 + \lvert \alpha_{11} \rvert^2}})\,,
\]
namely the normalised vector of $\alpha_{00} |0\rangle + \alpha_{01} |1\rangle$ ({\em resp.}\ $\alpha_{10} |0\rangle + \alpha_{11} |1\rangle$). \\
To translate measurements in Isabelle we first need a predicate {\em select\_index} such that {\em select\_index n i j} outputs true if the $j$th element of the computational basis has a $1$ at the $i$th spot of its label and false otherwise.

\begin{figure}[H]
	\includegraphics[scale=0.32]{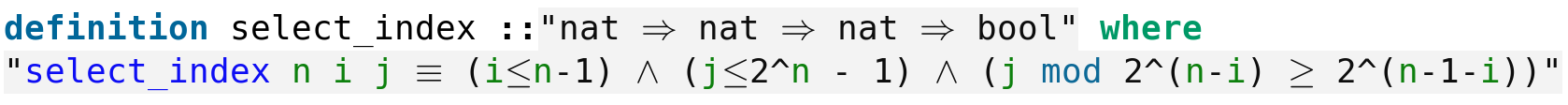}
	\caption{The {\em select\_index} predicate (see {\em Measurement.thy})}
\end{figure}
\noindent
Then given the state of a $n$-qubit system, we can compute the probability\footnote{We do not use any of the probability theory developed in Isabelle, we use ad hoc definitions instead. } of the outcome being $0$ ({\em resp.}\ $1$) when measuring the $i$th qubit.

\begin{figure}[H]
	\includegraphics[scale=0.32]{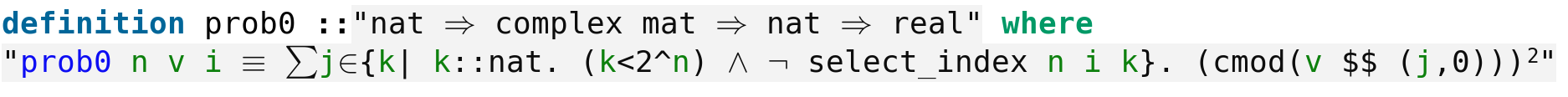}
	\includegraphics[scale=0.32]{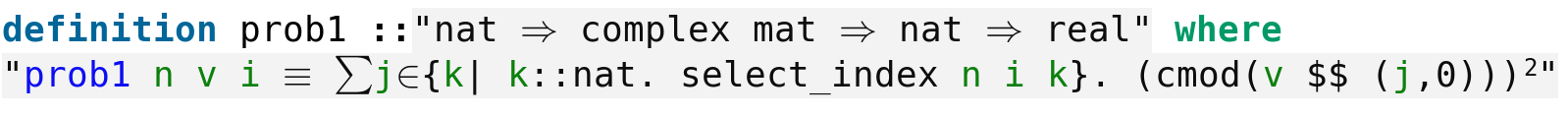}
	\caption{Computing the probabilities of outcomes (see {\em Measurement.thy})}
\end{figure}
\noindent
If the outcome of measuring the $i$th qubit is $0$ ({\em resp.}\ $1$), then {\em post\_meas0} ({\em resp.}\ {\em post\_meas1}) gives the new state of the system.

\begin{figure}[H]
	\includegraphics[scale=0.32]{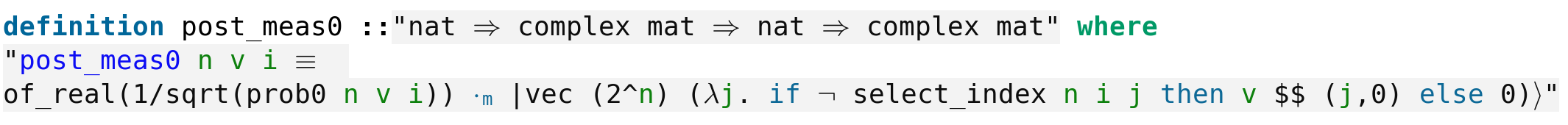}
	\includegraphics[scale=0.32]{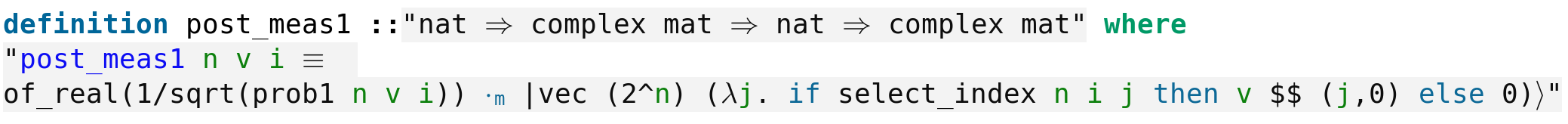}
	\caption{The new states of the system after outcome 0 and 1, respectively (see {\em Measurement.thy})}
\end{figure}
\noindent
Entanglement has some interesting consequences with respect to measurement. In quantum mechanics measurements of physical properties, such as momentum, position or spin, on entangled particles are perfectly correlated. In quantum computing this phenomenon can be illustrated through the Bell states. Given a Bell state, if one makes one measurement, then one gets either $0$ with probability $1/2$ or $1$ with probability $1/2$ whatever the qubit being measured (either the first or the second one). Moreover, in the case of two successive measurements of the first and second qubit, the outcomes are correlated. 
Indeed, in the case of $|\beta_{00}\rangle$ or $|\beta_{10}\rangle$ ({\em resp.}\ $|\beta_{01}\rangle$ or $|\beta_{11}\rangle$) if one measures the second qubit after a measurement of the first qubit (or the other way around) then one gets the same outcomes ({\em resp.}\ opposite outcomes), {\em i.e.}\ the probability of measuring $0$ for the second qubit after a measurement with outcome $0$ for the first qubit is $1$ ({\em resp.}\ $0$). 

\section{Theorems and Quantum Algorithms}
\label{sec:algo}

We present briefly the main theorems and algorithms formalized in the library. For a detailed presentation the reader is invited to consult a standard reference like \cite{Nielsen:2011:QCQ:1972505}. 

\subsection{The No-Cloning Theorem}

A notable theorem in quantum computation and quantum information is the so-called {\em no-cloning theorem} articulated by Wootters and Zurek \cite{Wootters:1982zz} and by Dieks \cite{Dieks82communicationby}. It is one of the earliest results in the field. Roughly, the no-cloning theorem states it is impossible to make an exact copy of an unknown quantum state. Since classical information can be copied exactly, this no-go theorem\footnote{In physics a no-go theorem states that a particular situation is physically impossible.} is one of the main differences between classical and quantum information. More precisely, given two non-orthogonal quantum states $|\phi\rangle$ and $|\psi\rangle$, there does not exist a quantum device that, when input with $|\phi\rangle$ ({\em resp.} $|\psi\rangle$), outputs $|\phi\rangle \otimes |\phi\rangle$ ({\em resp.} $|\psi\rangle \otimes |\psi\rangle$). First, we use Isabelle's locale mechanism to define a quantum machine. A quantum machine consists of a natural number $n$, a complex vector $s$, and a complex matrix $U$, plus the assumptions that $s$ has dimension $2^n$ and $U$ is a $2^{2n} \times 2^{2n}$ unitary matrix.

\begin{figure}[H]
	\includegraphics[scale=0.32]{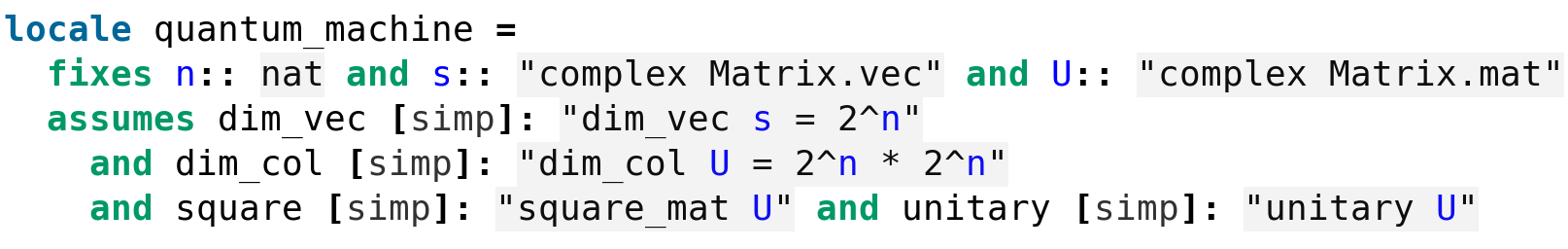}
	\caption{A quantum machine in Isabelle (see {\em No\_Cloning.thy})}
\end{figure}
\noindent
Second, we need to introduce the inner product $\langle v|w\rangle$ of two complex vectors $v,w$.

\begin{figure}[H]
	\includegraphics[scale=0.32]{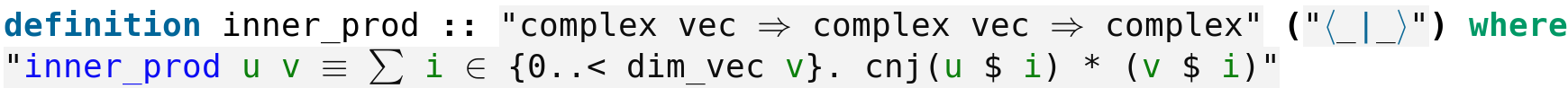}
	\caption{The inner product of two complex vectors (see {\em Quantum.thy})}
\end{figure}
\noindent
Recall that for every complex vector v one has $\lVert v\rVert^2 = \langle v|v\rangle$, and two complex vectors $v,w$ being orthogonal means their inner product $\langle v|w \rangle$ is $0$. 

\begin{figure}[H]
	\includegraphics[scale=0.32]{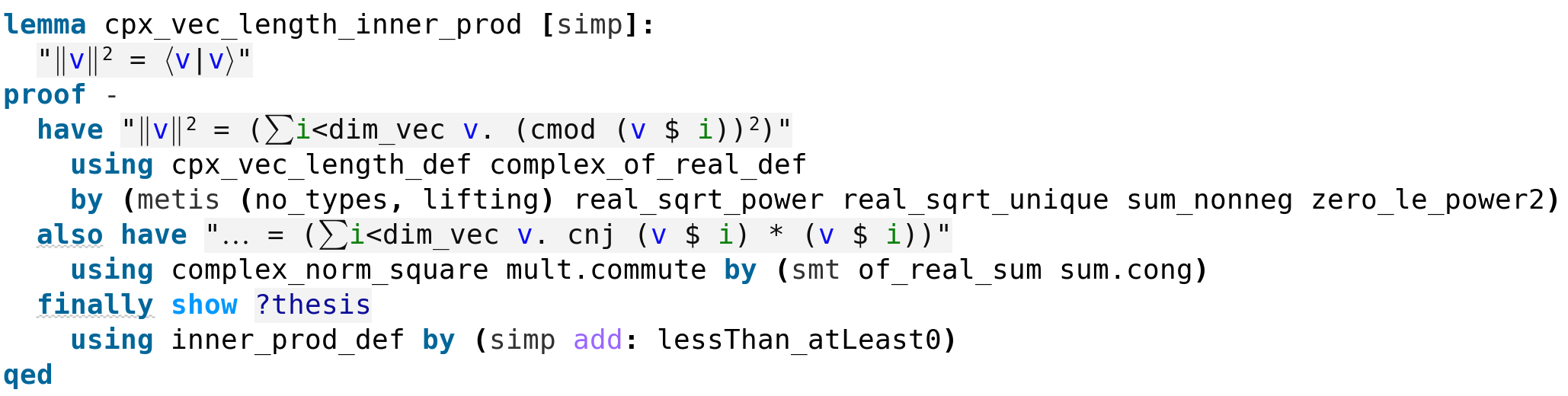}
	\caption{The squared length of a complex vector is equal to its inner product with itself (see {\em Quantum.thy})}
\end{figure}
\noindent
Thus, in Isabelle the no-cloning theorem is formalised as follows.

\begin{figure}[H]
	\includegraphics[scale=0.32]{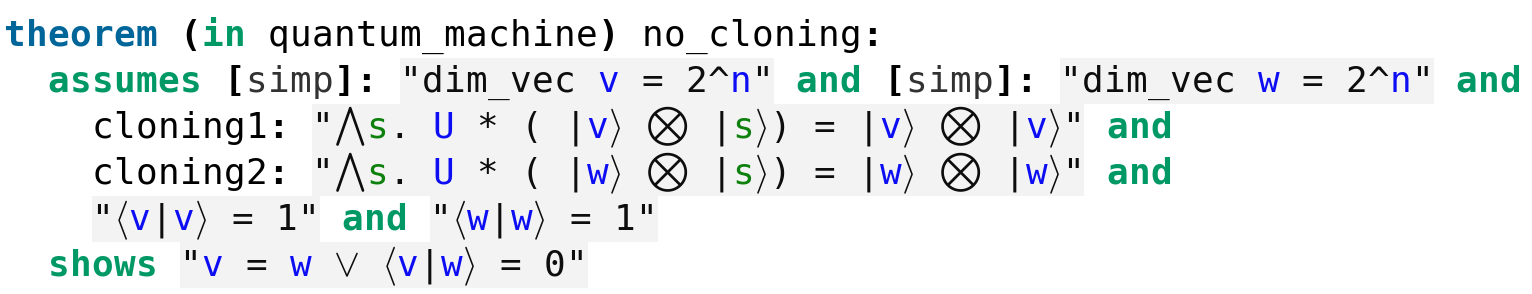}
	\caption{The no-cloning theorem (see {\em No\_Cloning.thy})}
\end{figure}
\noindent
In other words, if someone has built a quantum machine which is able to copy two quantum states ({\em i.e.}\ two normalised complex vectors), then these two states are either identical or orthogonal. The proof relies on the Cauchy-Schwarz inequality: 
\[
\lvert \langle v|w\rangle\rvert^2 \leq \langle v|v\rangle \langle w|w\rangle
\]
for every complex vectors $v$ and $w$. 

\begin{figure}[H]
	\includegraphics[scale=0.32]{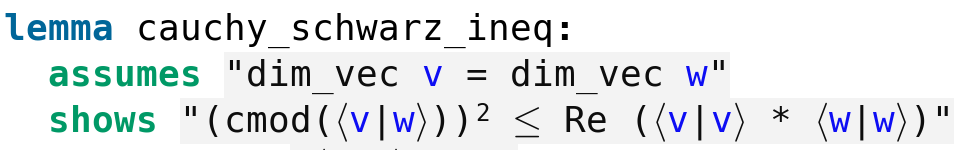}
	\caption{The Cauchy-Schwarz inequality (see {\em No\_Cloning.thy})}
\end{figure}
\noindent
In the snippet above one needs to take the real part of $\langle v|v\rangle \langle w|w\rangle$, since Isabelle is not able to notice immediately that $\langle v|v\rangle \langle w|w\rangle$ is a real number and so the real part is required for type-checking.

\subsection{Quantum Teleportation}

The quantum teleportation protocol has already been formalised with the proof assistant Coq \cite{Boender2015FormalizationOQ}. We follow closely this previous formalisation to give a counterpart in Isabelle. \\
First, we outline the protocol introduced in the seminal paper of Bennett {\em et al.}\ \cite{Bennett93teleportingan}. The quantum teleportation allows the transmission of an unknown quantum state  between a sender and a receiver in the absence of a quantum channel using only an entangled pair and a classical channel.  Let us assume that Alice in London wants to send Bob in Paris an unknown quantum state $|\varphi\rangle$. By sharing an EPR pair, each taking one qubit of the EPR pair, this feat can be achieved. Indeed, Alice can take the tensor product of $|\varphi\rangle$ with her half of the EPR pair to apply a cNOT gate (using $|\varphi\rangle$ as the control qubit) and then apply an Hadamard gate on $|\varphi\rangle$. Finally, she measures her two qubits, obtaining one of the four possible results: $00$, $01$, $10$ or $11$. She sends these two classical bits to Bob using the classical channel at her disposal. Depending on Alice's two bits, Bob performs one of four predetermined operations on his half of the EPR pair. More precisely, if Alice's two bits are $00$ ({\em resp.} $01$, $10$, $11$) then Bob applies the identity ({\em resp.} Pauli's $X$ gate, Pauli's $Z$ gate, Pauli's $X$ gate followed by Pauli's $Z$ gate). It can be shown that as a result Bob recovers $|\varphi\rangle$! \\
In the quantum circuit below the single lines denote qubits, the top two lines being Alice's qubits while the last one is Bob's qubit. The first gate represents a cNOT gate, $H$ denotes an Hadamard gate, the meters represent measurements, the double lines are classical channels carrying the classical bits $M1$ and $M2$ obtained after the measurements. This circuit gives a concise description of the protocol outlined above. 

\begin{figure}[H]
	\centering
	\includegraphics[scale=0.4]{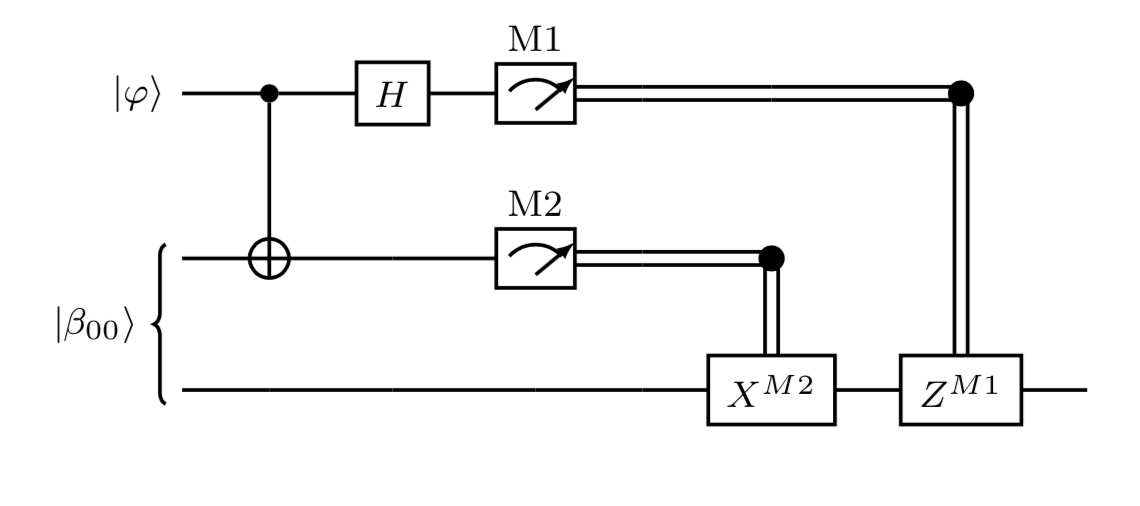}
	\caption{Circuit implementing the quantum teleportation protocol}
\end{figure}
\noindent
The formal specification of the protocol can be written in Isabelle as follows.

\begin{figure}[H]
	\includegraphics[scale=0.32]{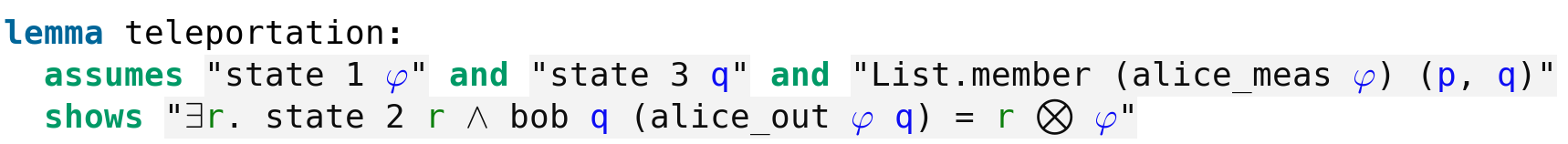}
	\caption{The quantum teleportation (see {\em Quantum\_Teleportation.thy}) \label{fig:teleportation}}
\end{figure}
\noindent
The function {\em alice\_out $\varphi$ q} outputs the two classical bits sent by Alice after the measurements. 

\begin{figure}[H]
	\includegraphics[scale=0.32]{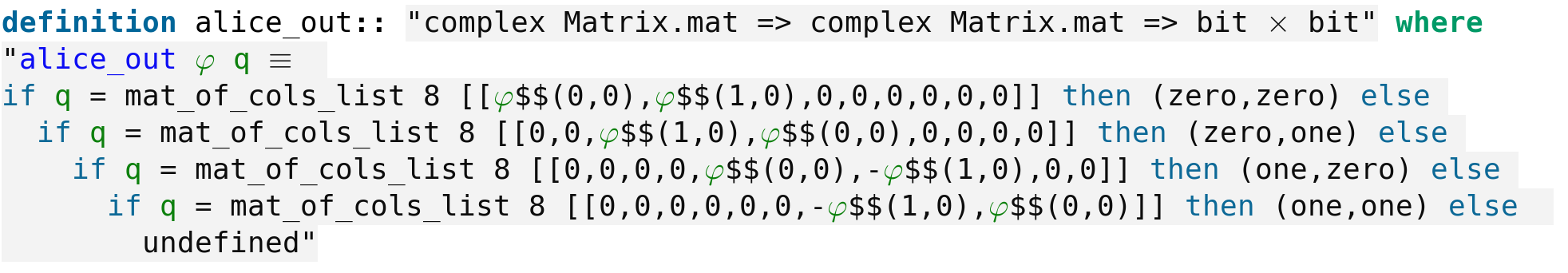}
\end{figure}
\noindent
The decoding function {\em bob q (alice\_out $\varphi$ q)} corresponds to the state of a 3-qubit system whose first and the second qubits are Alice's qubits after measurement and third qubit is Bob's qubit after performing his predetermined operation given the two classical bits sent by Alice.

\begin{figure}[H]
	\includegraphics[scale=0.32]{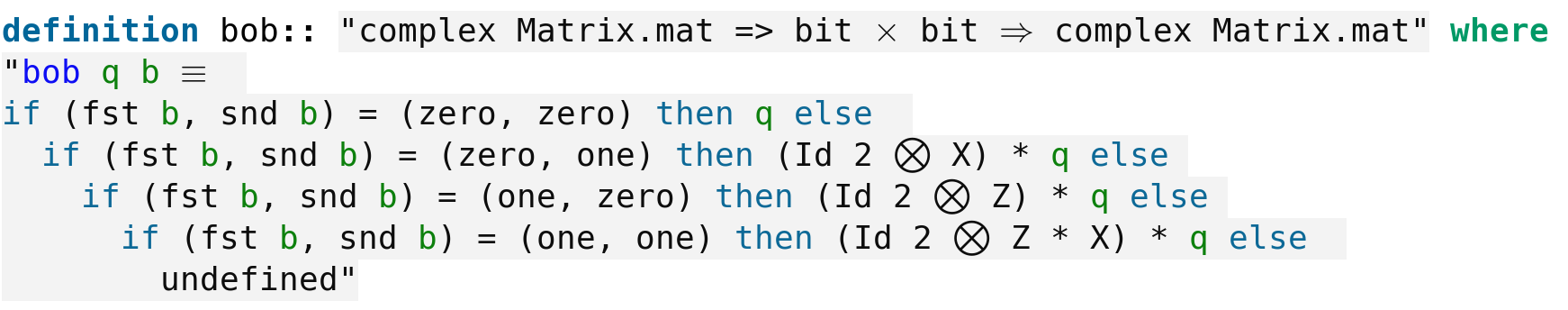}
	\caption{Bob's decoding function}
\end{figure}
\noindent
Then the formal specification \ref{fig:teleportation} asserts that the final state of Bob's qubit is nothing but $|\varphi\rangle$, namely the state given as argument and representing the unknown state Alice started with. The quantum state $|\varphi\rangle$ has been ``teleported'' from the first to the third position, {\em i.e.}\ from Alice to Bob. The existential quantification in the statement asserting that whatever Alice's two classical bits sent to Bob the state of the combined system always ``factors'' through $|\varphi\rangle$.

\subsection{The Deutsch-Jozsa  Algorithm} 

Deutsch in \cite{Deutsch85quantumtheory} was the first to demonstrate that a quantum computer could perform a task faster than any classical computer. His algorithm was improved later by numerous researchers. We explain below the purpose of Deutsch's algorithm. \\
A function taking values in $\lbrace 0,1\rbrace$ is {\em balanced} if it outputs $0$ for half of its inputs and $1$ for the other half. We start with a function $f:\lbrace 0,1\rbrace \rightarrow \lbrace 0,1\rbrace$. \\
Classically one needs two evaluations of $f$ to determine if the function $f$ is constant or balanced. Deutsch's quantum algorithm determines if  $f$ is constant or balanced using only one evaluation of $f$. This feat is made possible by quantum parallelism, {\em i.e.}\ the ability to evaluate a function $f(x)$ for many values of $x$ simultaneously. The quantum circuit implementing Deutsch's algorithm is drawn below.

\begin{figure}[H]
	\centering
	\includegraphics[scale=0.4]{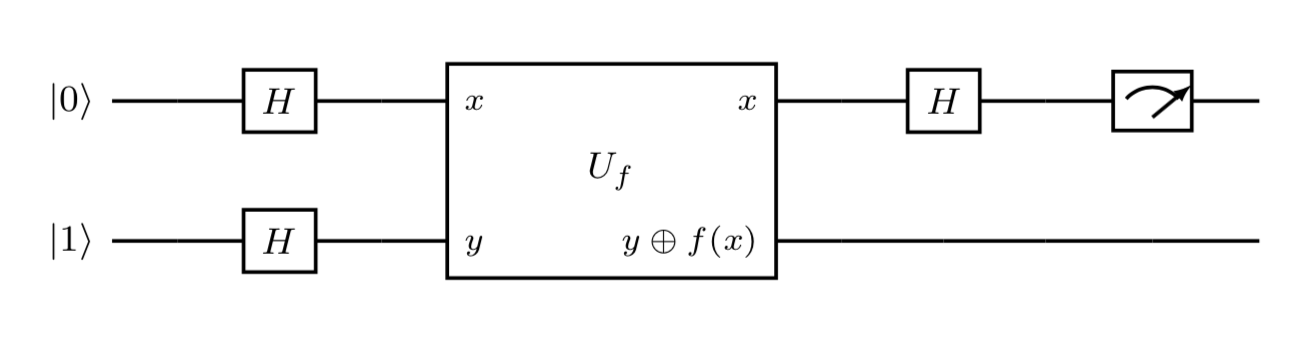}
	\caption{Circuit implementing Deutsch's algorithm}
\end{figure}
\noindent
Two qubits are prepared, one in the state $|0\rangle$ and another one in the state $|1\rangle$ .
A Hadamard gate is then applied to each of them followed by the unitary $U_f$. Afterward the second qubit remains unchanged while the first one is subject to another application of the Hadamard transform. Finally, the first qubit is measured. \\
In Isabelle the set-up is provided by the following locale. 

\begin{figure}[H]
	\includegraphics[scale=0.32]{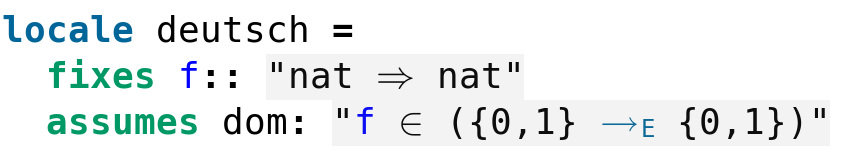}
\end{figure}
\noindent
Then we translate the algorithm in Isabelle, the last gate operation being translated by $H \otimes Id\,1$ since it leaves the second qubit untouched. Note that if time flows from left to right in the circuit, the code should be read from right to left, since the first matrix applied in a matrix multiplication is the one on the right.

\begin{figure}[H]
	\includegraphics[scale=0.32]{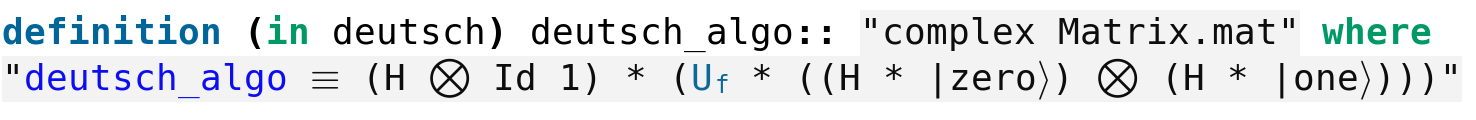}
	\caption{Deutsch's algorithm (see {\em Deutsch.thy})}
\end{figure}
\noindent
Finally, we check the correctness of the algorithm.

\begin{figure}[H]
	\includegraphics[scale=0.32]{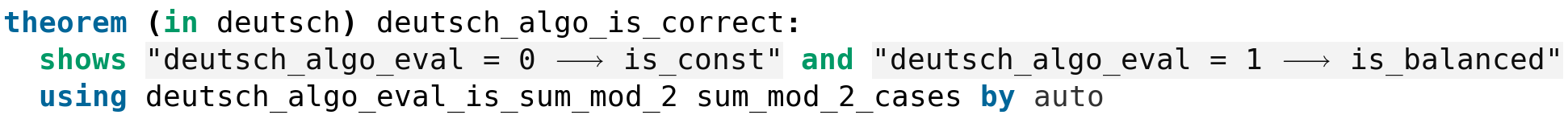}
	\caption{Deutsch's algorithm (see {\em Deutsch.thy})}
\end{figure}
\noindent
where {\em deutsch\_algo\_eval} is equal to $f(0) \oplus f(1)$, namely $f(0) + f(1)$ modulo 2.

\begin{figure}[H]
	\includegraphics[scale=0.32]{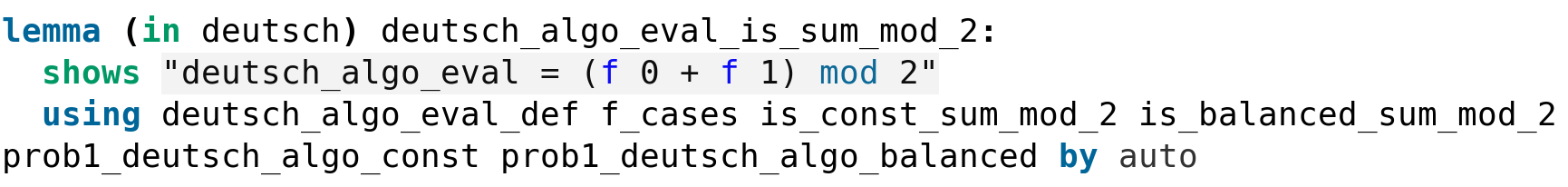}
\end{figure}
\noindent
Deutsch's algorithm has a generalisation, the so-called Deutsch-Jozsa algorithm, where the domain of $f$ has $2^n$ values. \\
Let us assume that we have a function $f:\lbrace 0,\dots, 2^n-1\rbrace \rightarrow \lbrace 0,1\rbrace$ which is either constant or balanced. In the following circuit for the Deutsch-Jozsa algorithm the wire annotated with $/^n$ represents a set of n qubits. For n = 1 one recovers the particular case of Deutsch's algorithm.

\begin{figure}[H]
	\centering
	\includegraphics[scale=0.4]{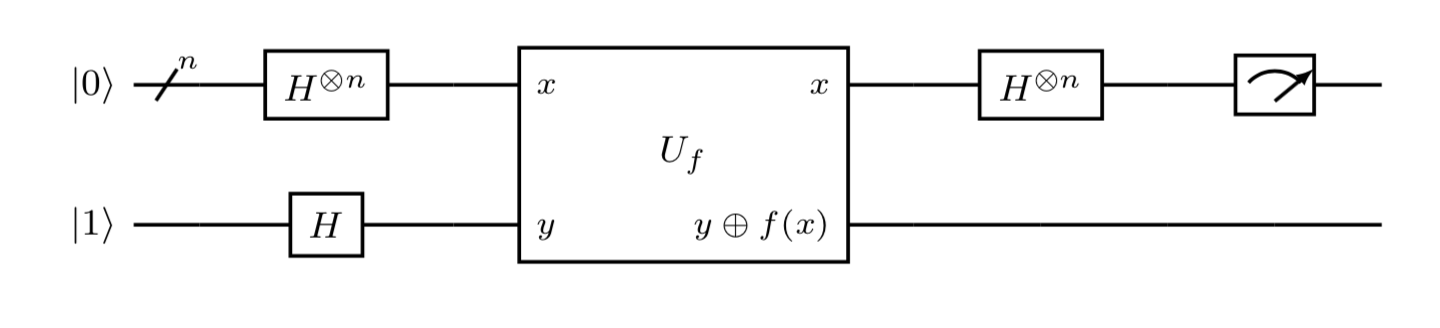}
	\caption{Circuit implementing the Deutsch-Jozsa algorithm}
\end{figure}
\noindent
The set-up in Isabelle is given by two locales where Bob promises Alice that he will use a function which is either constant or balanced.

\begin{figure}[H]
	\includegraphics[scale=0.32]{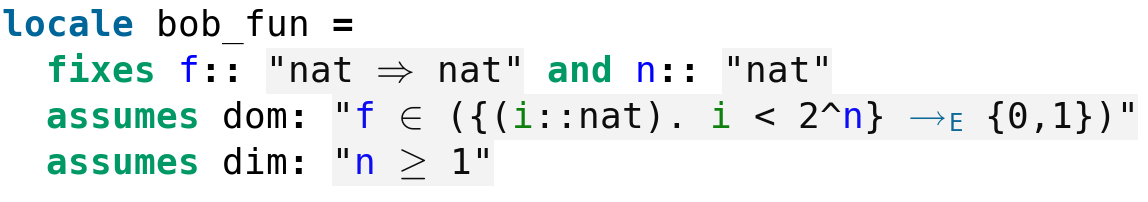} \\
	\includegraphics[scale=0.32]{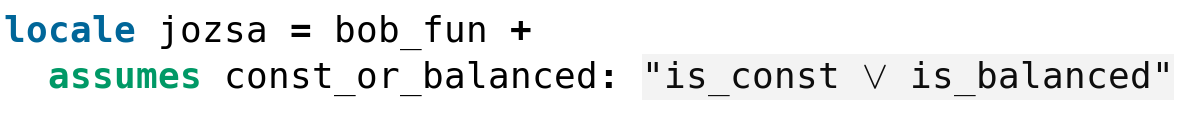}
\end{figure}
\noindent
Classically in the worst-case scenario Alice needs $\frac{2^n}{2} + 1$ queries to determine if Bob's function $f$ is constant or balanced. Indeed, Alice can get $\frac{2^n}{2}$ $0s$ before getting a $1$. However, using the Deutsch-Jozsa algorithm Alice can decide if $f$ is constant or balanced using only one evaluation of $f$. \\
The translation in Isabelle is similar to the one of Deutsch's algorithm except that the evaluation of the algorithm now requires the measurement of the first n qubits.

\begin{figure}[H]
	\includegraphics[scale=0.32]{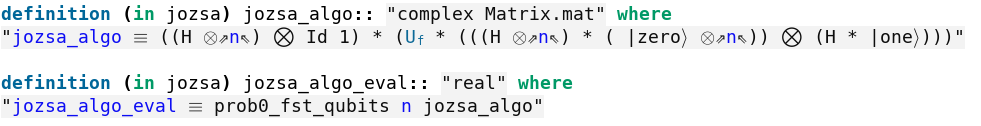}
	\caption{The Deutsch-Jozsa algorithm (see {\em Deutsch\_Jozsa.thy})}
\end{figure}
\noindent
Then one can certify the correctness of the Deutsch-Jozsa algorithm which outputs $1$ ({\em resp.} $0$) if and only if $f$ is constant ({\em resp.} balanced).

\begin{figure}[H]
	\includegraphics[scale=0.32]{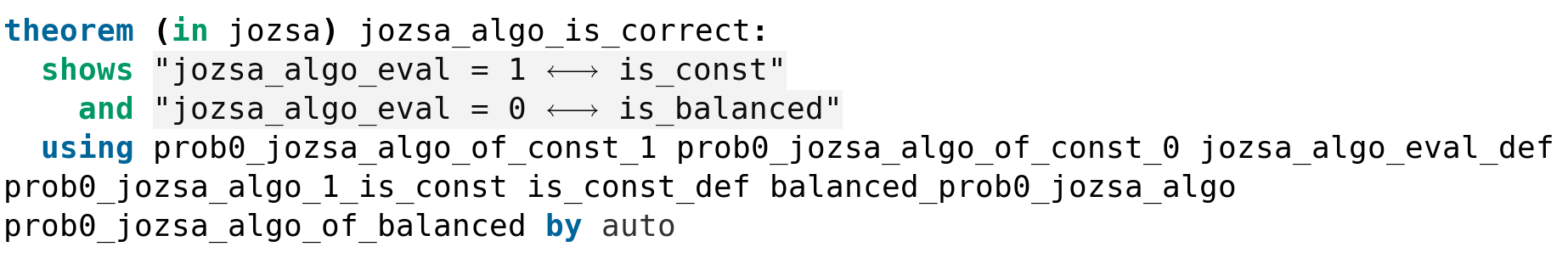}
	\caption{The Deutsch-Jozsa algorithm (see {\em Deutsch\_Jozsa.thy})}
\end{figure}

\subsection{The Quantum Prisoner's Dilemma}
\label{ssec:qPD}

We will assume that the reader is familiar with the Prisoner's Dilemma and the basic concepts of non-cooperative game theory \cite{RePEc:mtp:titles:0262650401}. The quantum version of the Prisoner's Dilemma was introduced by Eisert, Wilkens and Lewenstein in their classic article \cite{PhysRevLett.83.3077}. \\
The strategic space of the quantum game is given by the set of unitary $2\times 2$ matrices of the form

\[
\hat{U}(\theta,\varphi)=
\begin{pmatrix}
e^{i\varphi}\cos(\theta/2) & \sin(\theta/2) \\
-\sin(\theta/2) & e^{-i\varphi}\cos(\theta/2)	
\end{pmatrix}
\]
with $0 \leq \theta \leq \pi$ and $0 \leq \varphi \leq \pi/2$.
As noted in \cite{PhysRevLett.87.069801} the strategic space used by Eisert {\em et al.}\ consisting of these 2-parameter unitary matrices is only a subset of $SU(2)$ and as a consequence is unlikely to reflect any reasonable physical constraint. However, this subset already exhibits interesting properties arising in the quantum regime and as a consequence is worth studying. \\
The quantization  scheme is parametrized by a real $\gamma\in [0,\pi/2]$ which is a measure of the game's entanglement. For $\gamma = 0$ one recovers the classical game while $\gamma = \pi/2$ corresponds to a maximally entangled game. 

\begin{figure}[H]
	\includegraphics[scale=0.32]{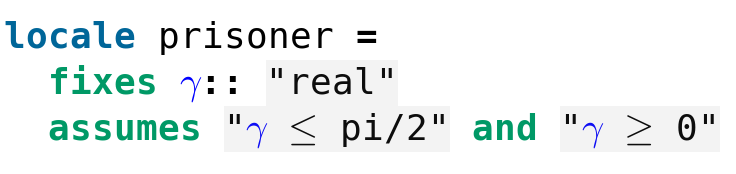} \\
	\includegraphics[scale=0.32]{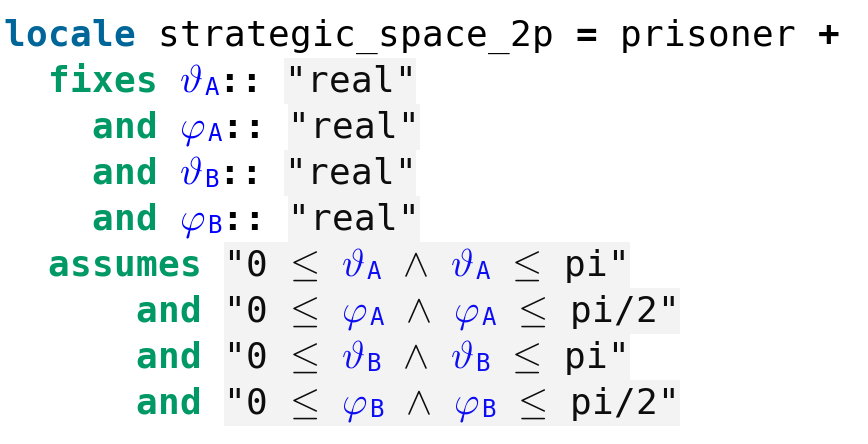}
	\caption{Our two players, Alice and Bob, and their parameters defining their strategies}
\end{figure}
\noindent
Then one defines a unitary operator $\hat{J}$ as $\hat{J}\coloneqq exp\lbrace i\,\gamma\;\hat{D}\otimes\hat{D}/2\rbrace$, where $\hat{D}\coloneqq \hat{U}(\pi,0)$ is the strategy to defect while $\hat{C}\coloneqq \hat{U}(0,0)$ is the strategy to cooperate.

\begin{figure}[H]
	\includegraphics[scale=0.32]{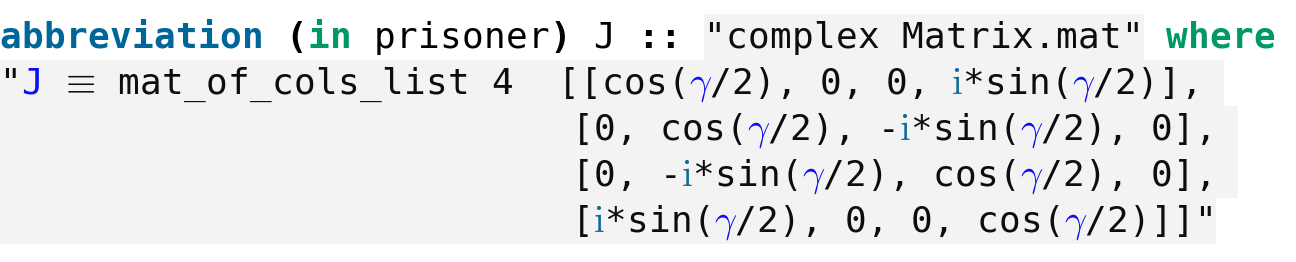}
\end{figure}
\noindent
If $\hat{U}_A\coloneqq \hat{U}(\theta_A, \varphi_A)$ ({\em resp.} $\hat{U}_B\coloneqq \hat{U}(\theta_B, \varphi_B)$) denotes Alice's ({\em resp.} Bob's) strategy, then the final state of the game is given by 
\[
|\psi_f\rangle \coloneqq \hat{J}^{\dagger} \, (\hat{U}_A \otimes \hat{U}_B) \, \hat{J} \, |00\rangle\,.
\]

\begin{figure}[H]
	\includegraphics[scale=0.32]{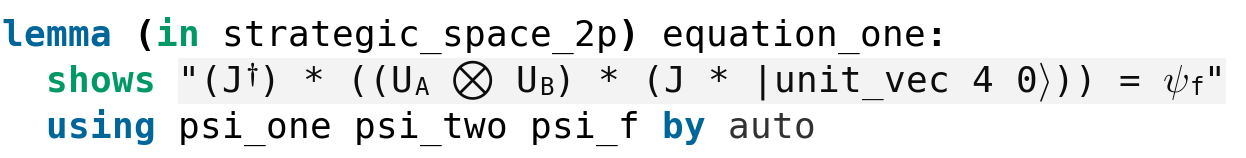}
	\caption{The final state of the game}
\end{figure}	
\noindent
Finally, Alice's expected payoff is calculated according to the following formula
\[
\$_A \coloneqq 3 P_{00} + P_{11} + 5 P_{10}\,, 
\]
while Bob's expected payoff is obtained by
\[
\$_B \coloneqq 3 P_{00} + P_{11} + 5P_{01}\,,
\] 
where $P_{xy} \coloneqq \lvert \langle xy|\psi_f\rangle\rvert^2$.

\begin{figure}[H]
	\includegraphics[scale=0.32]{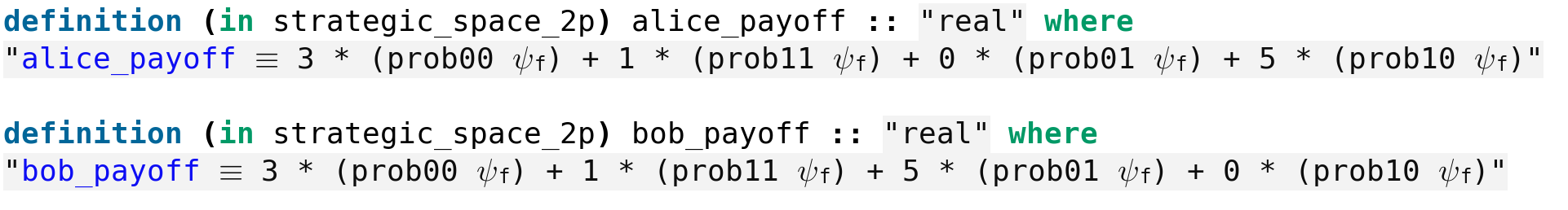}
	\caption{Alice's and Bob's expected payoffs}
\end{figure}
\noindent
To formalise in Isabelle the main results of \cite{PhysRevLett.83.3077}, we need to introduce formal definitions for Nash equilibriums and Pareto optimality in the context of our game and its restricted strategic space.

\begin{figure}[H]
	\includegraphics[scale=0.32]{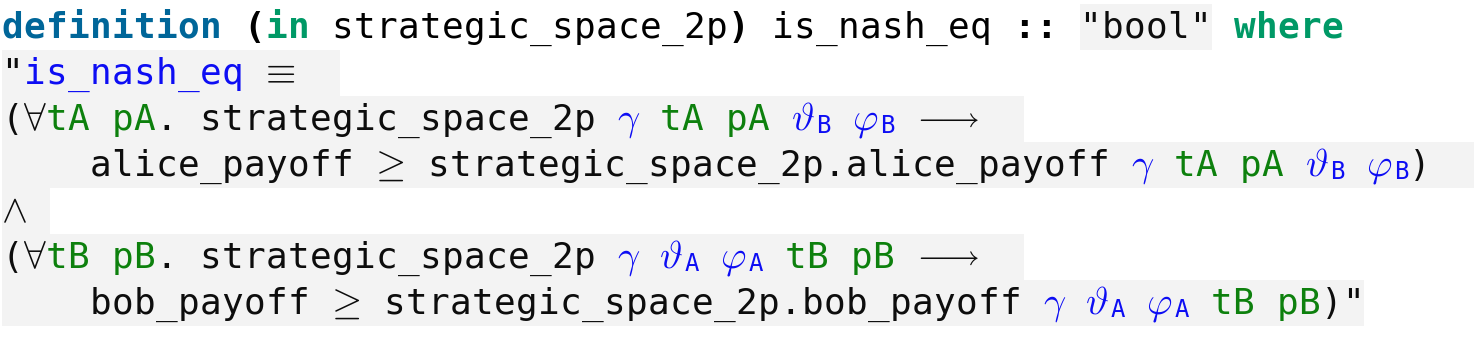}
	\includegraphics[scale=0.32]{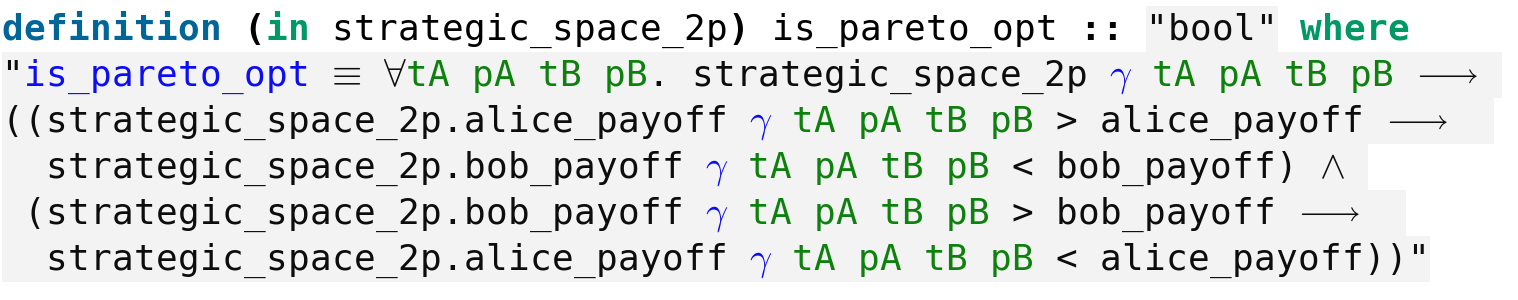}
	\caption{The formal definitions of Nash equilibrium and Pareto optimality, respectively}
\end{figure}
\noindent
In the classical game ($\gamma = 0$, also called the separable case) it is well known that both players defecting ({\em i.e.} playing the strategy $\hat{D}$, namely $\varphi_A = \varphi_B = 0$ and $\theta_A = \theta_B = \pi$) is a Nash equilibrium.

\begin{figure}[H]
	\includegraphics[scale=0.32]{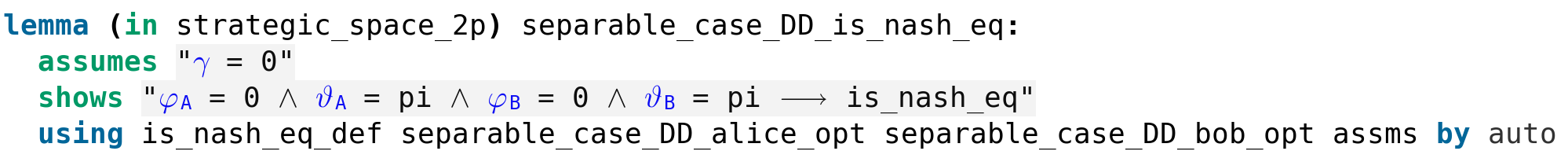}
\end{figure}
\noindent
First, the authors prove that in the maximally entangled quantum game ($\gamma = \pi/2$) both players defecting is no longer a Nash equilibrium. 

\begin{figure}[H]
	\includegraphics[scale=0.32]{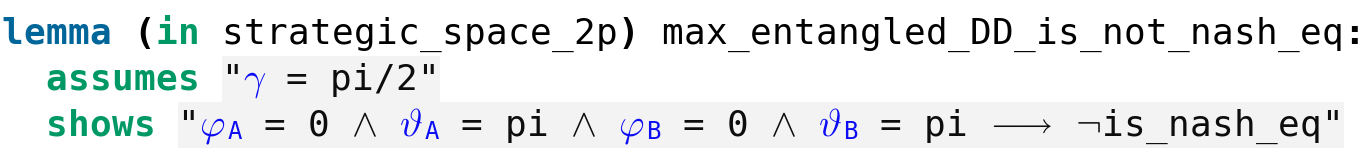}
\end{figure}
\noindent
Second, the authors introduce a new quantum strategy when $\gamma = \pi/2$, coined the {\em quantum move} and denoted $Q \coloneqq \hat{U}(0, \pi/2)$, with a high payoff (namely $3$) for both players resolving the prisoner's dilemma. They prove that both players playing $Q$ is a Nash equilibrium which is also Pareto optimal.

\begin{figure}[H]
	\includegraphics[scale=0.32]{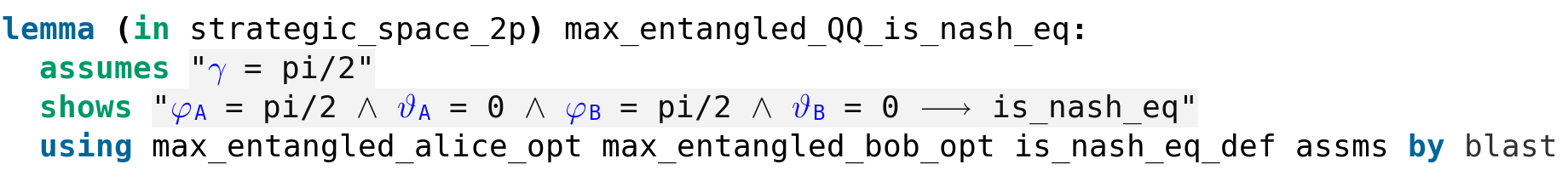}
	\includegraphics[scale=0.32]{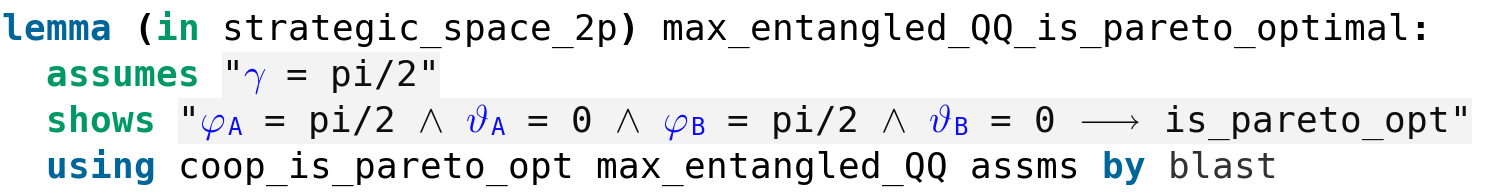}
	\caption{In the quantum regime a new Nash equilibrium appears which is Pareto optimal}
\end{figure}
\noindent
Finally, in the last part of their article Eisert {\em et al.}\ study an unfair version of the Prisoner's Dilemma where one player is restricted to classical strategies while the second player is not subject to such a restriction, {\em i.e.} Alice can play any strategy, either classical or quantum, while Bob can only play classical strategies. However, in the next section we point out a flaw in their treatment of the unfair version of the game. Indeed, we will see it is not true that the so-called {\em miracle move} as defined in \cite{PhysRevLett.83.3077} always gives quantum Alice a large reward against classical Bob and outperforms the so-called {\em tit-for-tat} strategy in an iterated game.

\section{The Unfair Version of the Quantum Prisoner's Dilemma}

Below we show the section in \cite{PhysRevLett.83.3077} on the quantum-classical version of the Prisoner's Dilemma, where Alice may use a quantum strategy while Bob is restricted to a classical strategy, is flawed. \\
In particular, the claim that the so-called miracle move, defined as $\hat{M}\coloneqq \hat{U}(\pi/2,\pi/2)$, gives Alice 
\begin{displayquote}
	at least reward $r = 3$ as pay-off, since $\$_A(\hat{M},\hat{U}(\theta,0)) \geq 3$ for any $\theta\in [0,\pi]$, leaving Bob with $\$_B(\hat{M},\hat{U}(\theta,0)) \leq \frac{1}{2}$ \cite[p.3079]{PhysRevLett.83.3077} 
\end{displayquote} 
is false. Indeed, for a maximally entangled game $\gamma = \frac{\pi}{2}$, for $\theta = \frac{\pi}{2}$ one has 
\[
\frac{1}{2} < \$_A(\hat{M},\hat{U}(\frac{\pi}{2},0)) = \$_B(\hat{M},\hat{U}(\frac{\pi}{2},0)) = 1 < 3\;.
\]
In the situation where Alice plays the miracle move while Bob is restricted only to classical strategies, for $0 \leq \gamma \leq \frac{\pi}{2}$ we have
\begin{align}
\$_A(\hat{M},\hat{U}(\theta,0)) & = \frac{1}{8}\, (21 + \cos(\gamma)^{2} (-3 + 14 \cos\theta) + 3 \sin(\gamma)^{2}-16 \sin\gamma\sin\theta) \\
\$_B(\hat{M},\hat{U}(\theta,0))  & = \frac{1}{8}\, (11 + \cos(\gamma)^{2} (7-6 \cos\theta)-7 \sin(\gamma)^{2} + 4 \sin\gamma\sin\theta)\;.
\end{align}
So, pluging $\gamma = \frac{\pi}{2}$ in equations (1) and (2) gives 
\[
\$_A(\hat{M},\hat{U}(\theta,0)) - \$_B(\hat{M},\hat{U}(\theta,0)) = \frac{5}{2}(1 -\sin\theta)
\]
admitting a minimum of $0$ when $\theta=\frac{\pi}{2}$. \\
In other words, the dilemma is not removed in favor of the quantum player contrary to the claim in \cite[III.C]{FAIntro} which reproduced the error in \cite{PhysRevLett.83.3077} supported by erroneous computations (the authors found $\$_A = 3 + 2\sin\theta$ and $\$_B = \frac{1}{2}(1 - \sin\theta)$ instead of $\$_A = 3 - 2\sin\theta$ and $\$_B = \frac{1}{2}(1 + \sin\theta)$).  \\
Indeed, Bob can immunize himself against Alice's miracle move by playing the {\em down-to-earth} move $\hat{E}$
\[
\hat{E}\equiv\hat{U}(\frac{\pi}{2},0) = \frac{1}{\sqrt{2}}
\begin{pmatrix}
1 & 1 \\
-1 & 1
\end{pmatrix}\;,
\]
the outcome being a draw $\$_A = \$_B = 1$. \\
Assuming $\gamma=\frac{\pi}{2}$, $\phi_B = 0$, we get the following pay-off matrix.

\begin{table}[H]
	\centering
	\setlength{\extrarowheight}{2pt}
	\begin{tabular}{cc|c|c|c|}
		& \multicolumn{1}{c}{} & \multicolumn{3}{c}{Bob} \\
		& \multicolumn{1}{c}{} & \multicolumn{1}{c}{$\hat{C}$}  & \multicolumn{1}{c}{$\hat{D}$}  & \multicolumn{1}{c}{$\hat{E}$} \\\cline{3-5}
		& $\hat{C}$ & $(3,3)$ & $(0,5)$ & $(\frac{3}{2},4)$ \\ \cline{3-5}
		Alice  & $\hat{D}$ & $(5,0)$ & $(1,1)$ & $(3,\frac{1}{2})$ \\ \cline{3-5}
		& $\hat{Q}$ & $(1,1)$ & $(5,0)$ & $(3,\frac{1}{2})$ \\ \cline{3-5}
		& $\hat{M}$ & $(3,\frac{1}{2})$ & $(3,\frac{1}{2})$ &$(1,1)$ \\ \cline{3-5}
	\end{tabular}
\end{table}
So, if Alice plays $\hat{M}$, the dominant strategy of Bob becomes $\hat{E}$, thereby doing substantially worse than if they would both cooperate, reproducing the dilemma. Moreover, nothing supports the claim that Alice 
\begin{displayquote}
may choose ``Always-$\hat{M}$'' as her preferred strategy in an iterated game. This certainly outperforms {\em tit-for-tat} [\dots] \cite[p.3079]{PhysRevLett.83.3077}.
\end{displayquote}
In conclusion, the ``miracle move'' as defined in \cite{PhysRevLett.83.3077} is of no advantage.



\section{Conclusions and Future Work}

Our work demonstrates that an extensive formalisation of quantum algorithms and quantum information theory in Isabelle/HOL is possible and not a fruitless exercise.
Indeed, the Letter \cite{PhysRevLett.83.3077} of Eisert {\em et al.}\ is a pioneering and highly cited article published in Physical Review Letters, a high-profile physics journal. The error uncovered therein is a notable unexpected outcome of our library. Indeed, this error had gone unnoticed in the field until our work and we found at least one subsequent published paper that reproduced it. After a private communication Eisert {\em et al.} acknowledged their error and they actually found a fix to re-establish their conclusions regarding what they call the ``miracle move''. An erratum was published by Physical Review Letters \cite{PhysRevLett.124.139901}. \\
Possible future applications of our library could include the verification of quantum cryptographic protocols, Isabelle having been successfully used in the past by Lawrence Paulson for the verification of cryptographic protocols using inductive definitions \cite{Paulson:1998:IAV:353677.353681}. A related work is the formalisation in Isabelle of parts of the Quantum Key Distribution algorithm by Florian Kamm\"uller using a framework extending attack trees to probabilistic reasoning on attacks \cite{KAMMULLER2019}. \\
Last, there is ongoing work in our library to formalise the quantum Fourier transform and unlock the potential formalisation of a wide range of more advanced quantum algorithms relying on it.

\begin{acknowledgements}
Anthony Bordg thanks Dr.\ Angeliki Koutsoukou-Argyraki, Dr.\ Wenda Li, Prof.\ Lawrence Paulson and Yiannos Stathopoulos for many discussions about Isabelle. Part of this work was carried while Yijun He and Hanna Lachnitt did an internship under the supervision of Dr.\ Anthony Bordg. This internship was hosted by Prof.\ Lawrence Paulson at the Department of Computer Science and Technology of the University of Cambridge and the authors thank him for his warm welcome and material support. During his internship Yijun He was supported by the ERC Advanced Grant ALEXANDRIA (Project 742178), funded by the European Research Council, and by the Cambridge Mathematics Placements (CMP) Programme. Hanna Lachnitt's internship was supported by an Eramus$+$ grant. 
\end{acknowledgements}

 \section*{Conflict of interest}

 The authors declare that they have no conflict of interest.

\end{document}